\documentclass[11pt]{article}
\usepackage{graphicx}
\usepackage{epsfig}
\usepackage{amssymb,amsfonts,amsmath,amsthm}
\usepackage{mathrsfs}
\usepackage{epstopdf}
\usepackage{hyperref,color}
\usepackage{setspace}

\pdfoutput=1

\begin{document}

\title{What becomes of a causal set}
\author{Christian W\"uthrich and Craig Callender\thanks{The authors contributed equally to this paper.}}
\date{30 January 2015}
\maketitle

\begin{abstract}\noindent
Unlike the relativity theory it seeks to replace, causal set theory has been interpreted to leave space for a substantive, though perhaps `localized', form of `becoming'. The possibility of fundamental becoming is nourished by the fact that the analogue of Stein's theorem from special relativity does not hold in causal set theory. Despite this, we find that in many ways, the debate concerning becoming parallels the well-rehearsed lines it follows in the domain of relativity. We present, however, some new twists and challenges. In particular, we show that a novel and exotic notion of becoming is compatible with causal sets. In contrast to the `localized' becoming considered compatible with the dynamics of causal set theory by its advocates, our novel kind of becoming, while not answering to the typical A-theoretic demands, is `global' and objective. 
\end{abstract}

\section{Introduction}

Contemporary physics is notoriously hostile to an A-theoretic metaphysics of time. A recent approach to quantum gravity promises to reverse that verdict: advocates of causal set theory (CST) have argued that their framework is consistent with a fundamental notion of `becoming'. A causal set, or `causet', is a discrete set of events partially ordered by a relation of causality.  The idea is that these sets  `grow' as new events are added one by one to the future of already existing ones; furthermore, this `birthing' process is said to unfold in a `generally covariant' manner and hence in a way that is perfectly compatible with relativity.  Here is Rafael Sorkin (2006) advertising the philosophical pay-off:

\begin{quote}
One often hears that the principle of general covariance [...] forces us to abandon `becoming' [...]. To this claim, the CSG dynamics provides a counterexample. It refutes the claim because it offers us an active process of growth in which `things really happen', but at the same time it honors general covariance. In doing so, it shows how the `Now' might be restored to physics without paying the price of a return to the absolute simultaneity of pre-relativistic days. 
\end{quote}
The claim is that CST, or at least CST augmented with a dynamics such as classical sequential growth (CSG) dynamics, rescues temporal becoming and our intuitive notion of time from relativity. The claim is routinely made in the pertinent physics literature, and has even found its way into popular science magazines.\footnote{Cf.\ e.g.\ Dowker (2003, 38).} One might not believe that our intuitive notion of time needs or deserves rescuing, but there is no denying that if this claim is correct it would have significant consequences for the philosophy of time. Specifically, it may underwrite a `growing block' model of the metaphysics of time, as John Earman (2008) has speculated. 

This paper has two main themes, one concerned with kinematics and the other with dynamics. After presenting the basics of CST in the next section, we investigate the possibilities for becoming in the theory's kinematics in Section 3.  We show that Stein's becoming theorem in relativity is false in CST and try, in vain, to use the resulting freedom to escape a dilemma imposed on becoming by relativity. Then, in Section 4, we turn to the CST dynamics in search of a more robust notion of becoming.  Trying to square this sense of becoming with (discrete) general covariance has many costs.  However, we show that if one is willing to pay them, a novel and exotic form of becoming is compatible with relativity. Conclusions follow in Section 5.

\section{The basics of CST}

The guiding idea of CST is that the fundamental structure of the world consists of a discrete set of elementary events partially ordered by a relation that is essentially causal. The theory finds its inspiration in a theorem (Malament 1977) that shows the precise sense in which the causal ordering of a sufficiently well-behaved relativistic spacetime determines its geometry, up to a conformal factor. Paraphrased roughly, the causal order of spacetime contains all the information we care about except for the local scale or `size' of spacetime. Motivated by this result, Sorkin and others formulated an approach to quantum gravity wherein discrete events supply the scale information and causal relations supply the rest.  Quantum versions are being developed, but we will focus on the classical theory. 

The basic structure of the theory is the {\em causet} $\mathcal{C}$, i.e., an ordered pair $\langle C, \preceq\rangle$ consisting of a set $C$ of otherwise featureless events and a relation `$\preceq$' on $C$ which satisfies the following conditions:
\begin{enumerate}
\item $\preceq$ induces a partial order on $C$, i.e., it is a reflexive, antisymmetric, and transitive relation;
\item  local finitude, i.e., $\forall x,z \in C, \mbox{card}(\{y \in C| x\preceq y \preceq z\}) <\infty$.\footnote{More colloquially, the cardinality of all these sets has to be less than $\aleph_0$.}
\end{enumerate}
These simple conditions constitute the basic kinematic assumptions of CST. The demanded antisymmetry entails that the structure cannot contain the causal-set equivalent of closed timelike curves. The local finitude of causets means that they are discrete structures, and this discreteness is what leads to some relevant differences concerning the issue of becoming in relativity. 

If what at larger scales looks like a relativistic spacetime fundamentally is a causal set, then causal set theory must give an account of how relativistic spacetimes emerge from causal sets. But just how do relativistic spacetimes emerge from causal sets? On the standard way of conceptualizing the problem, a necessary condition for the emergence of relativistic spacetimes from causal sets is that the spacetimes appropriately approximates the fundamental causal set at the scales at which it offers an adequate description. A classical spacetime $\langle \mathcal{M}, g_{ab}\rangle$ is said to {\em faithfully approximate} a causal set $\langle C, \preceq\rangle$ just in case there is an injective function $\phi: C \rightarrow \mathcal{M}$ such that 
\begin{enumerate}
\item the causal relations are preserved, i.e.\ $\forall x, y\in C, x \preceq y$ iff $\phi(x) \in J^-(\phi(y))$, where $J^-(X)$ designates the causal past of the set $X$ of events in $\mathcal{M}$;
\item on average, $\phi$ maps one element of $C$ onto each Planck-sized volume of $\langle \mathcal{M}, g_{ab}\rangle$ (Smolin 2006, 210);
\item and $\langle \mathcal{M}, g_{ab}\rangle$ does not have `structure' at scales below the mean point spacing.
\end{enumerate}
The first condition demands that the causet's causal relations are preserved on the emergent level of the relativistic spacetime. The second condition fixes the local scale. The third condition captures the idea that a discrete structure should not give rise to an emerging spacetime with significant curvature at a scale finer than that of the fundamental structure. Of course, this demand fixes the fundamental scale; in CST, it is common to fix it to the scale of Planck volumes. 

Such an injective function can easily be found if we simply let the co-domain of $\phi$ to be determined by a random `Poisson sprinkling' of events onto $\mathcal{M}$. Given such a co-domain of $\phi$ in $\mathcal{M}$, we obtain a causal set satisfying the kinematic axiom by lifting the set of events together with all the relations of causal precedence obtaining among these events. It is thus straightforward to find a causal set that is approximated by a given globally hyperbolic spacetime with bounded curvature. 

Note that it is important that the selected events in $\mathcal{M}$ are picked randomly and do not form some regular pattern such as a lattice. If the set of events exhibits too much regularity, then Lorentz symmetry would be broken because we would be able to distinguish, at an appropriately coarse-grained scale, such a lattice from its Lorentz boosted analogue. Since there can be no such observable differences, the selection of events must be sufficiently random and irregular (Dowker et al.\ 2004).

It turns out that most causal sets sanctioned merely by the kinematic axiom do not stand in a relation of faithful approximation to spacetimes with low-dimensional manifolds. Thus, the vast majority of causal sets are not approximated by a relativistic spacetime. This is the so-called `inverse problem' or `entropy crisis' of causal set theory (Smolin 2006). Only if this difficulty is successfully solved can causal set theory become a viable approach to finding a quantum theory of gravity. 

Most efforts in causal set theory are thus directed at this issue. An obvious strategy to address the problem is to identify further conditions that a causal set must satisfy in order to be an acceptable model of a fundamental structure. The idea, then, is to formulate some additional axioms that appropriately restrict the permissible causal sets to just include mostly only those that can be faithfully approximated by a relativistic spacetime. These additional principles are thought to specify a `dynamics' and thus to select the dynamical, and hence physical, models of the theory among its kinematic models that merely satisfy the basic axiom. Before we get to the dynamics, let us turn our attention to (kinematical) causal sets and the possibilities they offer for an advocate of becoming.

\section{Facing the same dilemma?}

Is temporal becoming compatible with special relativity? Howard Stein (1991) famously proved that there is a sense of becoming compatible with Minkowski spacetime.  The reason for this is simply that the events of Minkowski spacetime $\langle \mathbb{R}^4, \eta_{ab}\rangle$ can be partitioned into past, present, and future in a way that respects the geometric structure of Minkowski spacetime. However, as we have stressed (Callender 2000 and W\"uthrich 2013), just to identify {\em some} partition of events in Minkowski spacetime and to call it `becoming' does not entirely remove the pressure on becoming from special relativity. What is addtionally required is a reason to regard that choice of partition as answering the A-theorist's plea to identify in the fundamental layer of reality what in her view is required to ground our temporal experience.  Focusing on the present, the worry is that any identification of a present in special relativity either answers to the presentist's explanatory request or is compatible with the structure of Minkowski spacetime, {\em but not both}. For example, one might introduce a foliation of spacetime into spacelike hypersurfaces totally ordered by `time'. Presumably, that would complement a presentist notion of a (spatially extended) present and of becoming, but at the price of introducing structure not invariant under automorphisms of Minkowski spacetime and hence arguably violating special relativity. Conversely, the present can be identified with invariant structures such as a single event or the surface of an event's past lightcone, and successive presents as a set of  events on a worldline or as a set of past lightcones totally ordered by inclusion, respectively, but such structures will have radically different properties from those ordinarily attributed to the present by those seeking to save it (see Callender 2000 and W\"uthrich 2013)---e.g.\ if we take the past lightcone as the present, then the big bang counts, counterintuitively, as `now'. Does the advocate of becoming face a similar dilemma in the context of CST?

In order to address that question, let us see whether we can construct a `present' from the resources of CST. Beginning with the event of the `here-now', one very natural definition of the events co-present with the `here-now' are those events on a `spacelike slice', technically a `maximal antichain', i.e., a maximal set of events such that any two events are incomparable in terms of the relation $\preceq$. A sequence of presents would then be a partition of a causet into such maximal antichains. There are a number of problems with a present thus defined. First, maximal antichains, by definition, do not have any structure. If such a `spatially extended' present were to have any spatial structure at all, then this structure must somehow be induced by, and thus be ontologically dependent upon, the larger---`temporally extended'---structure of the causal set. Second, for any given event `here-now', there are in general many maximal antichains of which it is an element. Thus, in a loose analogy to the many ways in which Minkowski spacetime can be foliated, the present in the sense of the set of events co-present with the here-now would thus not be uniquely defined. Third, and relatedly, a partition of a causal set into such maximal antichains would not be invariant under automorphisms of its structure. Thus, it seems that a `spatially extended' present in a causal set would very much run into difficulties of the sort encountered in special relativity. 

Let's probe deeper. In special relativity, Stein's theorem tells us, essentially, that any binary relation `is definite as of' adapted to the structure of temporally oriented Minkowski spacetime must coincide with `is in the causal past of' lest it degenerate into the trivial or the universal relation, modulo a choice of temporal direction.\footnote{A universal binary relation obtains between any two objects in the domain, while a trivial binary relation only obtains between any object and itself. Since, given a domain, all trivial relations are extensionally identical (and hence extensionally identical to `identity', the most important trivial relation), as are all universal relations, we use the definite article in both cases.} One might now expect that a statement analogous to Stein's theorem holds in CST as well. After all, there is an obvious sense in which the causal structure of causal sets is very much like that of Minkowski spacetime---indeed, CST is premised upon the idea that special relativity gets the causal structure of spacetimes as partial orderings among events basically right. Thus, it may appear as if a causal set is merely a discrete version of Minkowski spacetime. This expectation, as natural as it may be, is disappointed. 

Stein's notion of becoming is expressed as a binary relation $R$ between spacetime events. The relation $R$ can be interpreted as `is settled as of', `having become as of', or `is determinate as of' and similar notions. Imposing various conditions on becoming, Stein proves that a non-boring relation that respects the basic structure of Minkowski spacetime exists.  It turns out to be the relation that obtains between and only between an event and events in that event's causal past.  In particular, he assumes that $R$ is a reflexive, non-trivial, and non-universal relation on a Minkowski spacetime $\langle\mathbb{R}^n, \eta_{ab}\rangle$ of at least two dimensions $(n\geq2)$ invariant under automorphisms that preserve the time-orientation and generally is Lorentz covariant. Stein then shows that if $Rab$ holds for some ordered pair of points $\langle a, b\rangle$, with $a, b \in \mathbb{R}^n$, such that $ab$ is a past-pointing (timelike or null) vector, then for any pair of points $\langle x, y\rangle$ in $\mathbb{R}^n$, $Rxy$ holds if and only if $xy$ is a past-pointing vector. The upshot is that for any event $p$ in Minkowski spacetime, all the events in $p$'s causal past have become for it. 

\begin{figure}
\centering
\epsfig{figure=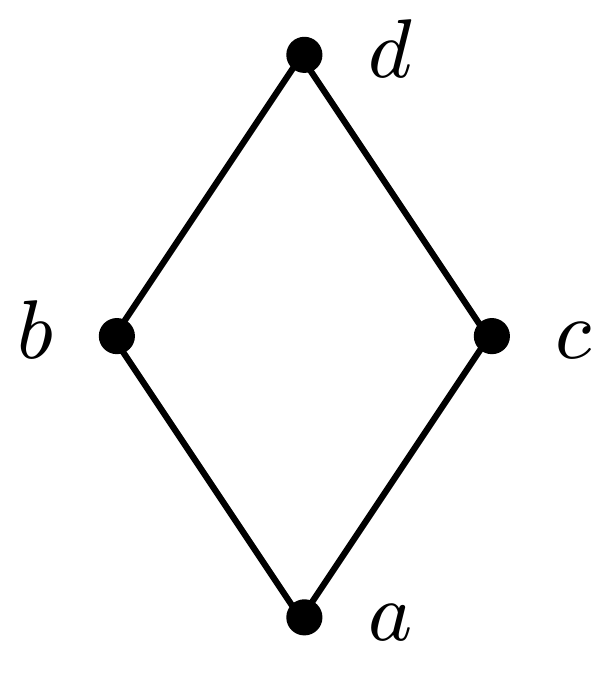,width=0.2\linewidth}
\caption{\label{fig:counterexample} The counterexample causet.}
\end{figure}

The analogue of this theorem, however,  is straightforwardly false in CST.   To quickly see this, consider the simple causet in Figure \ref{fig:counterexample}---the `counterexample causet'.\footnote{In the figure, we follow the usual practice of not including edges implied by reflexivity and transitivity.}
On this causet, a reflexive and transitive relation $R$ can be defined (set-theoretically, as is standard) as follows:
\begin{equation*}
R = \{\langle a,a\rangle, \langle b,b\rangle, \langle c,c\rangle, \langle d,d\rangle, \langle b,a\rangle, \langle c,a\rangle, \langle d,b\rangle, \langle d,c\rangle, \langle d,a\rangle, \langle b,c\rangle, \langle c,b\rangle\}.
\end{equation*}
In other words, this relation $R$---`is definite as of'---holds of any event and itself, of any event and any other event in its causal past, and of any `spacelike related' pairs of events such as $b$ and $c$ in Figure \ref{fig:counterexample}, and not otherwise. It is clear that this relation does not obtain only between events and events in their causal past, as it obtains between the spacelike related events $b$ and $c$!  Nor is this $R$ the trivial reflexive relation or the universal relation (e.g.\ $\langle b,d\rangle$ is not an element).  Nonetheless $R$ is invariant under automorphisms of the structure, as it only relies on the causal relations themselves, except for the last two pairs, which are however symmetrically included and obtain between points with identical `relational profiles' and hence are invariant under structure-preserving maps. In fact, we would expect to find relations violating the analogue of Stein's theorem whenever we have `non-Hegelian pairs' of events, i.e.\ pairs of events whose relational profile is identical. Automorphisms of a causal set map events to events in the causal set such that if a pair of events was standing in a relation of causal precedence prior to the mapping, their image will also do so.\footnote{Although the counterexample causet does admittedly not give rise to a relativistic spacetime, it could be a small proper part of a much larger causet that can be faithfully embedded into a relativistic spacetime. Toward the end of this section, we will return to the issue of how relevant and generic the counterexample is.} Of course, one may take this very fact to indicate that non-Hegelian pairs are not physically distinct events. 

Specifically for the counterexample causet, the pairs $\langle b,c\rangle$ and $\langle c,b\rangle$ can be included in $R$ without $R$ collapsing into the universal relation, as  happens in Minkowski spacetime. There, whenever a spacelike vector is included in $R$, invariance under automorphisms demands that {\em all} spacelike vectors are in $R$; requiring transitivity then collapses $R$ to the universal relation. This does not happen in cases such as the one represented in Figure \ref{fig:counterexample} because events with identical causal profile can be mapped onto each other without altering the causal structure at all. Thus, we see that in CST there can be non-trivial and non-universal relations of `being definite as of' that do not collapse to events in the causal past of the reference event. This gives the A-theorist novel ways of constructing physically kosher fundamental relations of co-presentness. But is it enough to drive a wedge into the dilemma faced by advocates of relativistic becoming?

Let us have a closer look at non-Hegelian subsets. A {\em non-Hegelian subset} $H\subseteq C$ of events in a causal set $\langle C, \preceq\rangle$ is a set consisting of distinct events $x_1,...,x_k$ in $C$ with the same relational profile, i.e., $\{x_1,...,x_k | \forall x_i, x_j, \forall z\in C \mbox{ such that } z \neq x_{i} \mbox{ and } z \neq x_{j}, \neg (x_i \preceq x_j) \mbox{ and } z\preceq x_i \leftrightarrow z\preceq x_j \mbox{ and } x_i\preceq z \leftrightarrow x_j \preceq z, \mbox{ where } i, j=1,...,k\}$.\footnote{Singleton sets of events are ruled out by the stipulation that any events in a non-Hegelian subset are pairwise unrelated by the ordering. Thus, $1<k\leq n$ for a causal set of $n$ events if there are any non-Hegelian subsets. The stipulation also implies that a non-Hegelian subset is an antichain, hence underwriting the following last sentence in the paragraph.} It is clear that any pair of distinct elements of a non-Hegelian subset cannot stand in the relation $\preceq$, i.e., they are by necessity `spacelike' to one another.

Now suppose we have a causal set with a non-Hegelian pair, i.e., a non-Hegelian subset of cardinality 2. The relation $R$ can thus symmetrically obtain between them in a way that leaves $R$ automorphically invariant, which is what led to the violation of Stein's result transposed to CST. If we subtract from $R$ all pairs which stand in $\preceq$, then we will end up with the non-Hegelian subsets.  Since their elements stand in $R$ symmetrically, these will be events which can be interpreted to be `determinate as of' one another. In this sense $R \;\setminus \preceq$ gives us an automorphically invariant way to define co-presentness. And this in a theory which is supposed to ground relativity and whose only fundamental relation is relativistic causal precedence!

Do we now have a tool allowing us to thwart the original dilemma between answering the presentist's explanatory needs and compatibility with relativity? For that to work, we would have to find large non-Hegelian subsets consisting of a nearly maximal antichain for the present to be at least almost global, and we would need many of them to have a decent sequence of subsequent presents. 

However, this is not what we generically find in causal sets.\footnote{`Generic' once the Kleitman-Rothschild hordes have been contained.} Although we know of almost no pertinent analytical results or numerical estimates, we suspect that large and many non-Hegelian subsets are few and far between.\footnote{An exception is an unpublished result found by David Meyer: the expected number of non-Hegelian pairs for a causet obtained from $N$ samples of a uniform process in an Alexandrov neighbourhood of $(1+1)$-dimensional Minkowski spacetime is 1 in the limit as $N$ goes to infinity. Since larger non-Hegelian subsets would contain more non-Hegelian pairs, their expected number would presumably be smaller and quickly tend to zero in the limit as the non-Hegelian subset grows.}  The reason is that there are many more irregular structures that satisfy the kinematic axiom than there are structures which are sufficiently regular to sustain large and many non-Hegelian subsets. If this is right, then there are no grounds on which to expect that the few and small remaining non-Hegelian subsets can generically give rise to any macroscopic present.

What about those causal sets which do have large and many non-Hegelian subsets and thus satisfy the condition necessary for an evasion of the dilemma? They clearly satisfy the kinematic axiom, so are not ruled out unless we impose additional dynamical laws that they violate. Yet observe that in these rare circumstances where the necessary condition is satisfied, we have regular `lattice' structures. As explained above, such highly regular structures would lead to a detectable violation of Lorentz symmetry. This has the great virtue of making it empirically testable whether---if CST is true at all, of course---the fundamental causal set exhibits such regularities. We take the absence of any empirical tests pointing to a violation of Lorentz symmetry (Mattingly 2005) to be an indication that the actual fundamental causal set---again, if any---cannot be highly regular. 

In sum, rather than escape the dilemma, it seems CST embraces it and even makes it rigorous.  That is, generically there will not be non-Hegelian subsets sufficient to express the present, and when there are, they will violate Lorentz invariance.

Of course, an advocate of becoming happy with Stein's relation can easily find a counterpart within causal sets. This would make becoming more local and observer-dependent, just as the Stein relations based on the causal structure of Minkowski spacetime are.\footnote{Clifton and Hogarth's (1995) similar view is an instance of observer-dependent becoming.} Given the structure of causets, it is straightforward to define relativistically kosher forms of becoming that essentially imitate the past lightcone becoming already compatible with the geometry of Minkowski spacetime; for instance, let `is definite as of' coincide with the causet order relation. However, to those seeking a genuinely `tensed' metaphysics of time, Stein's result has always had limited appeal.  At best it defines a notion of becoming compatible with Minkowski spacetime.  But if one desires that Minkowski spacetime {\em itself} grows or changes, as many metaphysicians of time do, then Stein's project is simply seen as irrelevant.  As Callender (2000) and Skow (2009, 668n) point out, there is a difference between notions of becoming and flow that are observer- or event-dependent and those that are independent of observers or events. If one wants the world to become, as tensers do, then one wants a more substantial perspective-independent sense of becoming.  Can CST provide us with this?

\section{Taking growth seriously}

There is nothing in the kinematics of CST that suggests any kind of ontological growth.  To find anything smacking of growth, one needs to turn to the dynamics, which is imposed to restrict the vast set of kinematically possible causets to the physically reasonable models of the theory. The usual dynamics for a causal set is a law of sequential growth. What grows are the number of elements, and it is assumed that the `birthing' of new elements is stochastic. Suppose $\Omega(n)$ is the set of $n$-element causets.  Then the dynamics specifies transition probabilities for moving from one $\mathcal{C} \in \Omega(n)$ to another $\mathcal{C}' \in \Omega(n+1)$.

Innumerable growth laws are possible.  Yet a remarkable theorem by Rideout and Sorkin (2000) shows that if the classical dynamics obeys some natural conditions such as label-independence and relativistic causality, then the dynamics is sharply constrained.  In particular, it must come from a class of dynamics of sequential growth known as `generalized percolation'.  Since the differences within this class will not matter for what follows, we can illustrate the idea with the simplest classical sequential growth dynamics that satisfies the Rideout-Sorkin theorem, namely, {\em transitive percolation}, a dynamics familiar in random graph theory. 

A simple way to understand this dynamics is to imagine an order of element births, labeling that order using integers $0, 1, 2...$ such that they are consistent with the causal order, i.e., if $x \preceq y$, then $\mbox{label}(x) < \mbox{label}(y)$.  (The reverse implication does not hold because the dynamics at some label time may birth a spacelike event, not one for which $x \preceq y$.) We begin with the causet's `big bang', the singleton set.  Now when element 2 is birthed, there are two possibilities: either it is causally related to 1 or not, i.e., $1 \preceq 2$ or $\neg(1 \preceq 2)$.  Transitive percolation assigns a probability $p$ to the two elements being causally linked and $1-p$ to the two elements not being causally linked. Ditto now for element 3, which has probability $p$ of being causally linked to 1 (2) and $1-p$ of not being causally linked to 1 (2). The dynamics enforces transitive closure, so if $1 \preceq 2$ and $2 \preceq 3$, then $1 \preceq 3$. Another way to conceive of the dynamics is that when each new causet $\mathcal{C}'$ is born, it chooses a previously existing causet $\mathcal{C}$ to be its ancestor with a certain probability.

The heart of the idea that CST rescues becoming involves taking sequential growth seriously: 
\begin{quote}
The phenomenological passage of time is taken to be a manifestation of this continuing growth of the causet. Thus, we do not think of the process as happening `in time' but rather as `constituting time'... (Rideout and Sorkin 2000, 024002-2)
\end{quote}
Becoming is embodied in the `birthing' of new elements. 

Although we are interested in becoming, we should immediately remark that sequential growth is certainly compatible with a tenseless or block picture of time. In mathematics a stochastic process is defined as a triad of a sample space, a sigma algebra on that space, and a probability measure whose domain is the sigma algebra.  Transition probabilities are viewed merely as the materials from which this triad is built.  In the case at hand, the sample space is the set $\Omega = \Omega(\infty)$ of past-finite and future-infinite labeled causets that have been `run to infinity'. The `dynamics' is given by the probability measure constructed from the transition probabilities; for details, see Brightwell et al (2003).  On this picture, the theory consists simply of a space of tenseless histories with a probability measure over them.\footnote{This interpretation corresponds to Huggett's first option (2014, 16), which is fully B-theoretic. When we consider `taking growth seriously', we mean to essentially follow the second route he offers: augmenting the causal structure with an additional, but gauge-invariant, dynamics.}

However, let's take the growth seriously.  There are different extents to which this can be done. At a more modest level, and consistent with explicit pronouncements by advocates of causet becoming, we can articulate a localized, observer-dependent form of becoming. Here, the idea is that becoming occurs not in an objective, global manner, but instead with respect to an observer situated within the world that becomes. The only facts of the matter concerning becoming are local, and are experienced by individual observers as they inch toward the future. In Sorkin's words, which are worth quoting in full, 

\begin{quote}
[o]ur `now' is (approximately) local and if we ask whether a distant event spacelike to us has or has not happened yet, this question lacks intuitive sense. But the `opponents of becoming' seem not to content themselves with the experience of a `situated observer'. They want to imagine themselves as a `super observer', who would take in all of existence at a glance. The supposition of such an observer {\em would} lead to a distinguished `slicing' of the causet, contradicting the principle that such a slicing lacks objective meaning (`covariance'). (2007, 158) 
\end{quote}

According to Sorkin, instead of ``super observers'', we have an ``asynchronous multiplicity of `nows' ''. It seems fairly straightforward that a perfectly analogous kind of becoming can be had in the context of Minkowski spacetime. Indeed, `past lightcone becoming', based on Stein's theorem, and `worldline becoming', as articulated by Clifton and Hogarth (1995), both satisfy the bill.\footnote{Cf.\ also Arageorgis (2012) who makes a similar point.} 

Although Sorkin himself remains uncommitted concerning whether the analogy holds, Fay Dowker (2014) rejects it, arguing that `asynchronous becoming' is not compatible with general relativity, but only with CST with a dynamics like the one provided by the classical sequential growth (and hence also not with the purely kinematic CST). The reason for this seems to be ultimately metaphysical, because only with the dynamics do we get not just the events, but their `occurrence'. Since in general relativity spacetime events do not `occur', goes the thought, there is no genuine form of becoming possible. Against this, we note firstly that (a large subsector of) general relativity certainly can be described in a `dynamical' manner via its many `3+1' formulations.\footnote{Cf.\ Wald (1984, Ch.~10).}  To make her objection, Dowker would first need to elaborate the reasons why a 3+1 dynamics does not provide the `occurrence' she desires.  Furthermore, we note here a possible tension.  If occurrence is simply a label for some events from the perspective of other events, then there is no problem---but then we note that such labels can be given consistently in general relativity too.  But if occurrence implies something metaphysically meaty, such as existence or determinateness---then there is a possible tension between occurrence and the local becoming envisioned by Sorkin and Dowker.  If spacetime events that are spacelike related do not exist for each other, for instance, then that is a radical fragmentation of reality.\footnote{In Pooley's view (2013, 358n), dynamical CST should best be interpreted as a ``non-standard $A$ Theory'' in Fine's (2005) sense, i.e., as giving up ``the idea that there are absolute facts of the matter about the way the world is.'' (2013, 334)}  Not only would that be a high cost to introduce becoming, but it is also one that, again, could be introduced in the ordinary theory.

Our present interest is to determine whether a more ambitious, objective, global, observer-independent form of becoming is compatible with CST-cum-dynamics in a way that does not violate the strictures of relativity. In other words, does Sorkin's assertion in the last sentence of the indented quote above holds up to scrutiny? We will argue that it does not and that there is a weak sense in which a fully objective kind of becoming with relativistic credentials can be had. 

Even before worrying about relativity, one might be concerned that an analogue of Jack Smart's `how fast does time fly?' objection applies when we turn to the dynamics (Smart 1966). Smart famously argued that if time changes and change is the having of different properties at different times, then it seems that at least two times are needed for any metaphysics wherein the present moves.  That seems to be the case here too. Remember that the elements being created are spatiotemporal.  What does a dynamics over variables that are spatiotemporal even mean?  We have an external time given by the dynamics---the time in which growth happens---and an internal time given by the spatiotemporal metric the causet inherits from its embedding into a relativistic spacetime. The causal set counterpart of Smart's question beckons: how fast are elements born?

Is Smart's objection fatal to the idea of cosmological `growth'?  Here philosophical opinion divides.  Anticipating Smart's question, C.D.~Broad (1938) argued that the kind of change that time undergoes is a {\em sui generis} kind of process.  It is not to be analyzed as qualitative change, i.e.\ the change of properties with respect to time.  It is its own thing.  We get a hint of that answer in Rideout and Sorkin's claim that birthing  {\em constitutes} time and is not {\em in} time.  The causet growth is time, in some sense, not something that happens in time.  Like Broad, Bradford Skow (2009) believes that a second-order time is not required to make sense of a substantive notion of temporal passage.  He regards this apparent second time dimension as a kind of metaphor to understand the action of primitive tense operators. For philosophers such as Broad and Skow, Smart's objection has no purchase.  Others, however, might complain that appeal to {\em sui generis} processes and primitive logico-linguistic devices leaves a lot to be desired in terms of physical clarity.\footnote{Or simply not evade the problem, as pressed by Pooley (2013, Section 4).}  No matter our personal reactions to this issue, we will bracket this worry since stopping at this point would be needlessly controversial.  After all, we are trying to give CST becoming its best chance.

The problem with taking the primitive growth as vindicating becoming is that advocates of CST uniformly wish to treat the labeling time as `fictitious'.  The reason is that the choice of label is tantamount to picking a time coordinate $x_0$ in a relativistic spacetime.  Any dynamics distinguishing a particular label order will be non-relativistic.  Not wanting the dynamics to distinguish a particular label (`coordinatization'), the authors impose {\em discrete general covariance} on the dynamics.  This is a form of label invariance.  The idea is that the probability of any particular causet arising should be independent of the path to get to that causet.  In particular, if $\alpha$ is one path from the singleton causet to an $n$-element causet, and $\beta$ is another path to the same causet, then the product of the transition probabilities along the links of $\alpha$ is the same as that for $\beta$ (and any other such path). 

To get a feel for this, suppose that the singleton set births a timelike related element, Alice's birthday, at label time $l=1$, and then this 2-element causet births a third element, Bob's birthday, spacelike related to the other two events at label time $l=2$.  That is path $\alpha$.  Path $\beta$ instead births Bob's birthday spacelike related to the singleton set, and then births Alice's birthday timelike related only to the singleton set. Discrete general covariance implies that the product of the transition probabilities getting from the singleton to that 3-element causet is the same.  Used as a condition to derive the dynamics, all sequential growth dynamics compatible with CST possess this symmetry. The further interpretation is that the probabilities respect this symmetry because the labels are pure gauge, that there is no fact of the matter about which path was taken.

\begin{figure}
\centering
\epsfig{figure=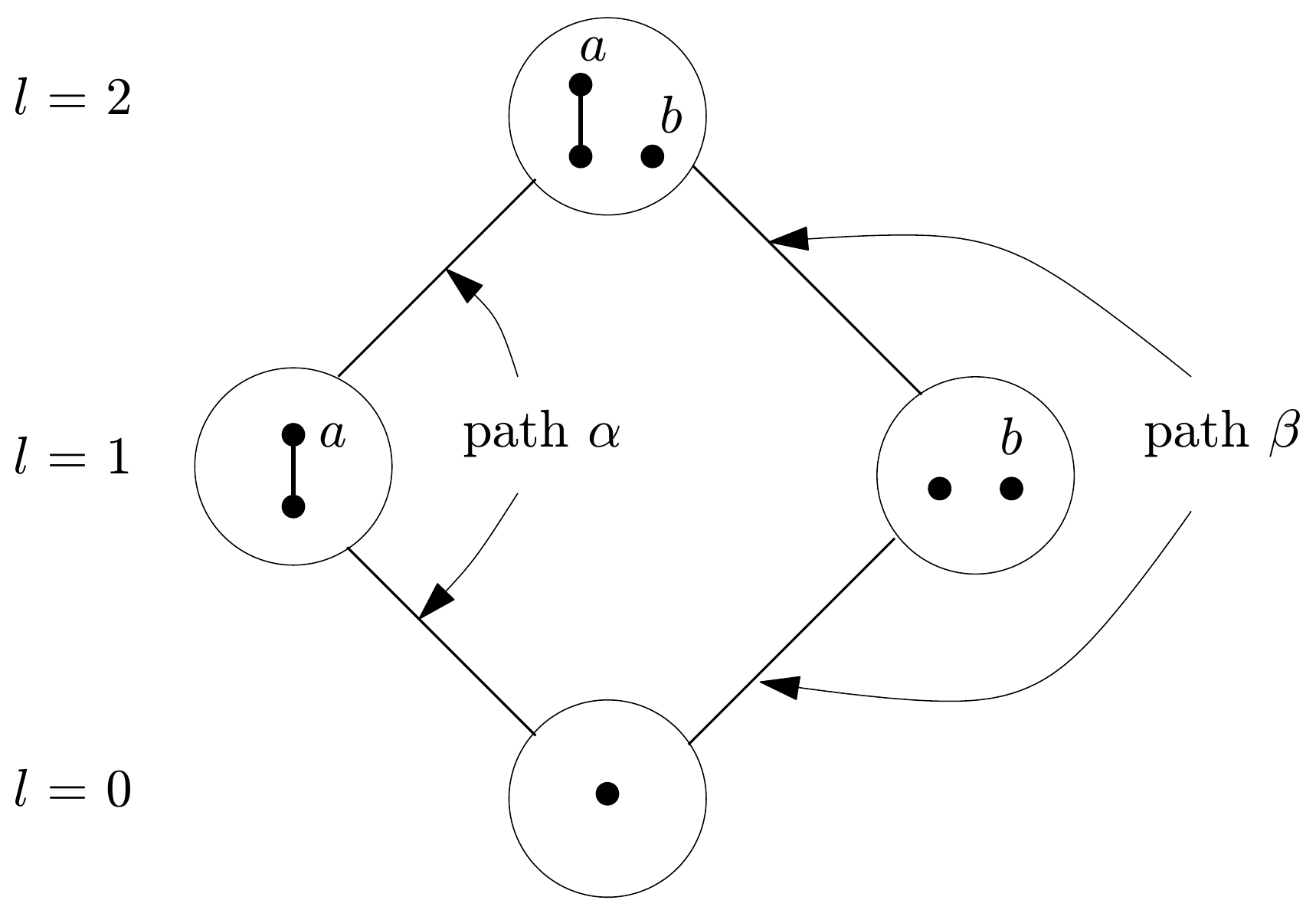,width=0.6\linewidth}
\caption{\label{fig:alicebob} Alice's and Bob's birthday parties come into being.}
\end{figure}

With this simple example in mind, one can immediately see the trouble with regarding this growth as a real physical process (see Figure \ref{fig:alicebob}).  Suppose the event $a$ timelike related to the singleton set is Alice's birthday party and suppose the event $b$ spacelike related to both is Bob's birthday party.  To enforce consistency with relativity, there is no fact of the matter about which one happened right after the singleton element event.  To say which one happened `first' is to invoke non-relativistic concepts.  It is therefore hard to understand how there can be growth happening in time. Seeing the difficulty here, Earman (2008) suggests a kind of philosophical addition to causal sets, one where we imagine that `actuality' does take one path or another.  With such a hidden variable moving up the causet, we do regain a notion of becoming.  But as Aristidis Arageorgis (2012) rightly points out, such a move really flies in the face of the normal interpretation of these labels as pure gauge.\footnote{Cf.\ also Butterfield (2007, 859f).} The natural suggestion, espoused by (almost all?) philosophers of physics, is then that the above tenseless interpretation is best because it does not ask us to imagine that one event came first.

Perhaps the sensible reaction to this problem is to abandon the hope that CST does produce a novel sense of becoming.  Still, we are tempted to press on.  The intuition motivating us is as follows.  True, the dynamics is written in terms of a choice of label, but we know that a consistent gauge invariant dynamics exists `beneath' this dynamics.  In fact, rewriting the theory in terms of a probability measure space, as indicated above, one can quotient out under relabellings to arrive at a label-invariant measure space (for construction and details, see Brightwell et al 2003).  And one thing that we know is gauge invariant is the number of elements in any causet.  Focusing just on these and ignoring any labeling, we do have transitions from $\mathcal{C}$ to $\mathcal{C}'$ and so on.  There is gauge-invariant growth.  

The problem is that we are generally prohibited from saying exactly what elements exist at any stage of growth.  Take the case of Alice and Bob above.  The world grows from $\mathcal{C}_1$ to $\mathcal{C}_2$ to $\mathcal{C}_3$. That's gauge invariant.  We just cannot say---not due to ignorance, but because there is no fact of the matter---whether $\mathcal{C}_2$ consists of the singleton plus Alice's party or the singleton plus Bob's party.  Causal set reality does not contain this information. There simply is no determinate fact as to whether $\mathcal{C}_2$ contains $a$ or $b$; but there is a determinate fact that it contains one of them. If it is coherent, therefore, to speak of a causet having a certain number of elements but without saying what those elements are, then CST does permit a new kind of---admittedly radical and bizarre---temporal becoming. 

Whether this notion of becoming is coherent depends on the identity conditions one has for events.  If to be an event, one has to be a particular type of event with a certain character, then perhaps the idea is not coherent. After all, what is the $\mathcal{C}_2$ world like?  It does not have Alice {\em and} Bob in it (that's $\mathcal{C}_3$), nor does it have {\em neither} Alice nor Bob in it (that's $\mathcal{C}_1$).  The world determinately has Alice or Bob in it, but it does not have determinately Alice or determinately Bob.  `Determinately' cannot penetrate inside the disjunction.  Notice that this feature is a hallmark of vagueness or of metaphysical indeterminacy more generally. Without going into any details of the vast literature on vagueness, let us note that there is a lively dispute over whether there can be ontological vagueness. The causal set program, interpreted as we have here, supplies a possible model of a world that is ontologically vague.  Further discussion of this model seems to us worthwhile.

First, we would simply like to point out that Ted Sider (2003) has supplied arguments that existence cannot be vague. That existence cannot be vague or indeterminate was a central assumption of his argument to four-dimensionalism in his [2001]. In fact, he asserts (2003, 135) that anyone who accepts the premise that existence cannot be vague is committed to four-dimensionalism, the thesis that objects persist by having temporal parts. To the extent to which many advocates of becoming reject four-dimensionalism anyway, they would thus be open to embrace ontological indeterminacy even if Sider's arguments of 2001 and 2003 were successful. And they may well not be: one of them, for instance, infers to the impossibility of vague existence from the claim that it cannot be vague how many things there are in a finite world (2001, 136f). Obviously, a defender of observer-independent becoming in CST may agree that it is at no moment vague how many events there exist, but nevertheless disagree that existence cannot be vague. Thus, we may have ontological indeterminacy without vagueness in the cardinality of the (finite) set of all existing objects. 

\begin{figure}
\centering
\epsfig{figure=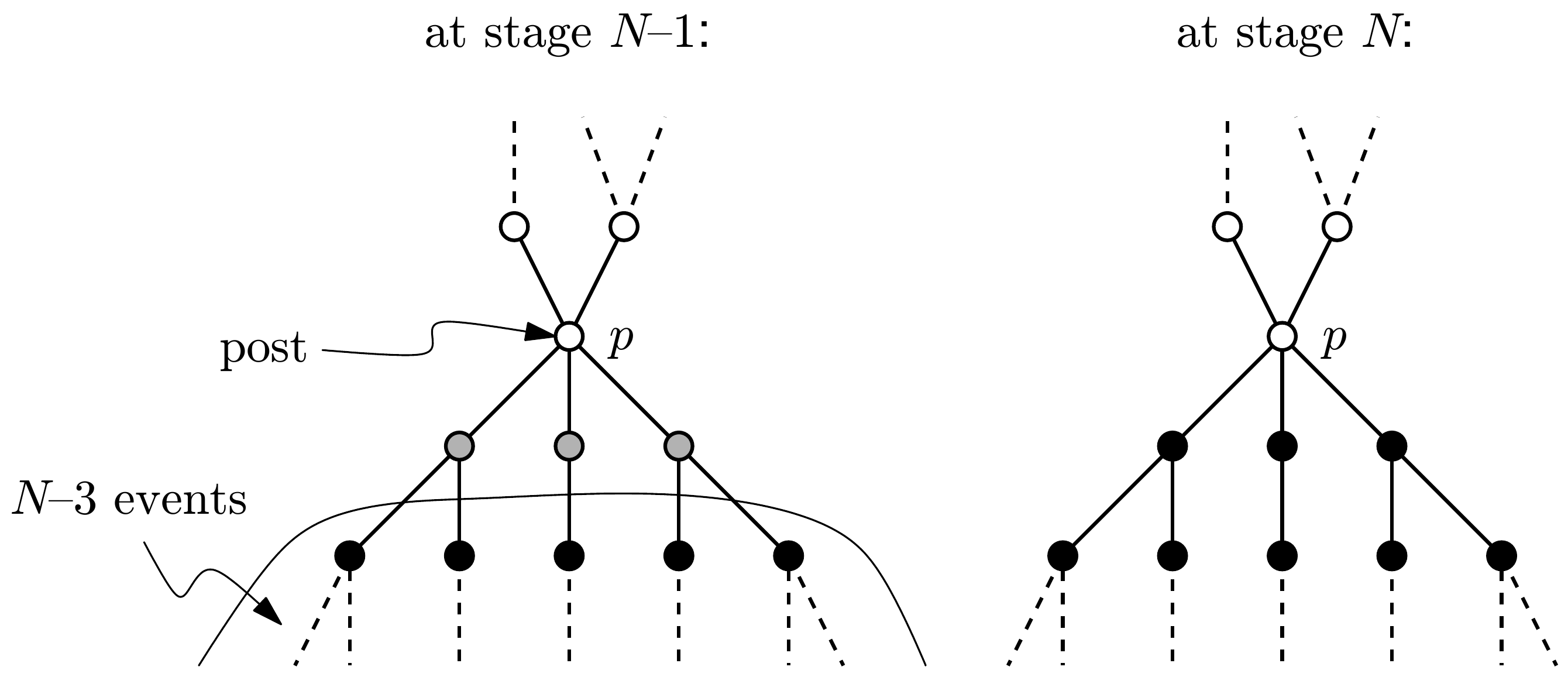,width=0.78\linewidth}
\caption{\label{fig:postgrowth} Becoming at post $p$.}
\end{figure}

One may be worried that on this notion of becoming in CST, no event in a future-infinite causet may ever be determinate until future infinity is reached, at which point everything snaps into determinate existence. This worry is particularly pressing as realistic causets are often taken to be future-infinite. So does any event ever get determinate at any finite stage of becoming? In general, yes. One way to see this is by way of example. As it turns out, causets based on transitive percolation in general have many `posts', where a {\em post} is an event that is comparable to every other event, i.e., an event that either is causally preceded or causally precedes every other event in the causet. Rideout and Sorkin interpret the resulting cosmological model as one in which ``the universe cycles endlessly through phases of expansion, stasis, and contraction [...] back down to a single element.'' (1999, 024002-4)\footnote{Cf.\ also Bollob\'as and Brightwell (1997).} Consider the situation as depicted in Figure \ref{fig:postgrowth}.
There is a post, $p$, such that $N$ events causally precede $p$, while all the others---potentially infinitely many---are causally preceded by $p$. At stage $N-1$, shown on the left, there exist $N-1$ events. At this stage, all the `ancestors' of $p$ except those three events which immediately precede $p$, shown in black, must have determinately come to be. Of the three immediate predecessors, shown in grey to indicate their indeterminate status, two must exist; however, it is indeterminate which two of the three exist. At the prior stage $N-2$, the grey set of events existing indeterminately would have extended one `generation' further back, as it could be that two comparable events are the last ones to come to be before the post becomes. At the next stage, stage $N$, $N$ events exist and it is determinate that all ancestors of $p$ exist. There is no ontological indeterminacy at this stage. Event $p$ has not yet come to be at either stage and is thus shown in white. At stage $N+1$, not shown in Figure \ref{fig:postgrowth}, event $p$ determinately comes into existence. At stage $N+2$, one of the two immediate successor to $p$ exists, but it is indeterminate which one. And so on. 

One may object that this interpretation of the dynamics of a future-infinite causet presupposes a given final state toward which the causet evolves. Even though everything in the preceding paragraph is true under the supposition that the final causet is the one represented in Figure \ref{fig:postgrowth}, the objection goes, at stage $N$ it is not yet determined {\em that} $p$ is a post, as there could have been other events spacelike-related to $p$. Given that it is thus indeterminate whether $p$ is indeed a post, and since this is the case for all events at finite stages, no events can thus snap into determinate existence at any finite stage of the dynamical growth process. 

First, it should be noted that even if this objection succeeds, it is still the case that it is objectively and determinately the case that at each stage, one event comes into being and that thus the cardinality of the sum total of existence grows. Although the ontological indeterminacy remains maximal, there is a weak sense in which there is objective, observer-independent becoming. Second, if the causet does indeed not `tend' to some particular future-infinite causet, then all existence would always be altogether indeterminate (except for the cardinality). There would be no fact of the matter, ever, i.e., at any finite stage, of how the future will be, or indeed of how anything ever is. If this is the right way to think about the metaphysics of the dynamics of CST, we are left with a wildly indeterminate picture. Third, it should be noted that the mathematics of the dynamics is only well-defined in the infinite limit; in particular, for there to be a well-defined probability measure on $\Omega$, we must take $\Omega = \Omega(\infty)$ (Sorkin 2007, 160n; Arageorgis 2012, Section 3), which can be interpreted to mean that the future-infinite `end state' is metaphysically prior to the stochastic dynamics that grows the causet to that `state'. 

Note the strange features of this metaphysics.  First, note that many philosophers, from Aristotle to today, have thought that the future is indeterminate (see, e.g., \O hrstr\o m and Hasle 2011 and references therein).  According to some versions of this view, it is determinately true that tomorrow's coin flip will result in either a head or a tail, but it is not determinate yet which result obtains.  Vagueness infects the future.  We note that the above causal set vagueness is quite similar, but with one big difference:  on the causal set picture, the past too can be indeterminate!  In our toy causal set, it is not true at $\mathcal{C}_3$ that $\mathcal{C}_2$ determinately is one way rather than the other.  

Second, as a causet grows, events that were once spacelike to the causet might acquire timelike links to future events.  If we regard the growth of a new timelike link to a spacelike event as making the spacelike event determinate, modulo the above type of vagueness, then this is a way future becoming can make events past.  That is, there is a literal sense in which one can say that ``the past isn't what it used to be''. Strange as this may sound, it should be noted that the `growth' of the past stands in perfect analogy to that in past-lightcone becoming. 

Finally, although we don't have space to discuss it here, observe that despite appearances transitive percolation is perfectly time reversal invariant.  This allows the construction of an even more exotic temporal metaphysics.  If we relax the assumption that events can only be born to the future of existing events, then it is possible to have percolation---and hence becoming---going both to the future and past.  Choose a here-now as the original point.  Then it is possible to modify the theory so that the world becomes in both directions, future and past. Of course, similarly, we could have a causal set that is future-finite and only grows into the past, and thus is past-infinite.

\section{Conclusion}

We have investigated the claim that CST rescues temporal becoming.  At the kinematical level, CST does offer new twists in dealing with time and relativity, but the basic contours of the relativistic challenge remains.  Serious constraints also threaten becoming if we take the time in CST's dynamics seriously too.  Here, however, if one is open to the costs of a sufficiently radical metaphysics, we maintain that there is a novel and exotic type of temporal becoming possible.

\section*{Acknowledgements}

We are indebted to Caro Brighouse, Juliusz Doboszewski, Sam Fletcher, Nick Huggett, David Meyer, Oliver Pooley, David Rideout, and Sebastian Speitel, as well as audiences at the Pacific APA in San Diego, the BSPS in Cambridge, and the Minnesota Center for Philosophy of Science for discussions and comments.


\vspace{1cm}
\begin{flushright}
Christian W\"uthrich\\
(as of 1 August 2015:) D\'epartement de philosophie, Universit\'e de Gen\`eve, Geneva, Switzerland\\
christian.wuthrich@unige.ch\\

\vspace{5mm}
Craig Callender\\
Department of Philosophy, University of California, San Diego, La Jolla, CA, USA\\
ccallender@ucsd.edu
\end{flushright}

\end{document}